\documentstyle[12pt]{article}
\begin{document}

\title{Current tensor with heavy photon for double hard photon emission
by longitudinally polarized electron}
\author{M.Konchatnij, N.P.Merenkov}
\date{}
\maketitle
\begin{center}

\small{\it{Kharkov Institute of Physics and Technology}}
\end{center}
\vspace{0.5cm}
\begin{abstract}
The electron current tensor for the scattering of heavy photon on
longitudinally polarized electron accompanied with two hard real photon
is considered. The contribution of collinear and semicollinear kinematics
is computed. The obtained result allows to calculate the corresponding
contribution into second order radiative correction to DIS or
electron--positron annihilation cross--sections with the next--to--leading
accuracy.
\end{abstract}

1.The recent polarised experiments on deep inelastic scattering [1,2]
cover the kinematical region $y\simeq0.9$, where the electromagnetic
corrections to the cross--section are extremely large . The first order
QED correction have been computed in [3,4], and it is of the order the
Born cross--section in this region. That is why the calculation of the
second order QED correction becames very important for interpretation of
these experiments. The first step in such calculation was done in [5]
where one--loop corrected Compton tensor with a heavy photon had been
considered. That is one of the contribution into polarized electron
current tensor which appears in the second order of perturbation theory.
Another contributions arise due to double hard photon emission and pair
production.

Here we calculate the contribution into the polarized electron current
tensor caused by two hard photon emission. We investigate double collinear
kinematics and semicollinear one. This allows us to compute the
corresponding second order radiative correction to different observables
with the next--to--leading accuracy in the same manner as it was done, for
example, for small--angle Bhabha scattering [6] and tagged photon
cross--sections at DIS [7] and electron--positron annihilation [8].

In the Born approximation the electron current tensor with longitudinally
polarized electron has a form
\begin{equation}\label{1Born}
L_{\mu\nu}^B = Q_{\mu\nu} + i\lambda E_{\mu\nu}\ , \ \ Q_{\mu\nu} =
-4(p_1p_2)g_{\mu\nu} + p_{1\mu}p_{2\nu} + p_{1\nu}p_{2\mu}\ ,
\end{equation}
$$E_{\mu\nu} = 4\epsilon_{\mu\nu\rho\sigma}p_{1\rho}p_{2\sigma} \ ,$$
where $p_1(p_2)$ is the 4--momentum of the initial (final) electron, and
$\lambda = 1 (-1)$ if the initial electron is polarized along (against)
its 3--momentum direction.

In the case of single collinear photon emission the corresponding
contribution into electron current tensor conserves the born structure for
radiation along the scattered electron momentum direction
\begin{equation}\label{2 single scattered photon}
L_{\mu\nu}^{(1)f} =
\frac{\alpha}{2\pi}\biggl[\frac{1+(1+y)^2}{y}\widetilde L_0 -
\frac{2(1+y)}{y}\biggr]dyL_{\mu\nu}^B \ , \ \
y=\frac{\omega}{\varepsilon_2} \ , \ \ \widetilde L_0 =
\ln\frac{\varepsilon_2^2\theta_0^2}{m^2} \ , \end{equation} and gets an
additional part (which is proportional to $i\lambda E{\mu\nu}$) for
radiation along the initial electron momentum direction [9]
\begin{equation}\label{3 single initial photon}
L_{\mu\nu}^{(1)i} =
\frac{\alpha}{2\pi}\biggl\{\biggl[\frac{1+(1-x)^2}{x}L_0 -
\frac{2(1-x}{x}\biggr]L_{\mu\nu}^B -2xi\lambda E_{\mu\nu}\biggr\}dx \ ,
\  \ \ x=\frac{\omega}{\varepsilon_1}
\ , \ \  L_0 = \ln\frac{\varepsilon_1^2\theta_0^2}{m^2} \ .
\end{equation}
In Eqs.(2) and (3) $\omega$ is the photon energy, $\varepsilon_1
(\varepsilon_2)$ is the energy of initial (final) electron, $m$ is the
electron mass, and parameter $\theta_0$ defines the angular phase space of
hard colinear photon. Index $i(f)$ labels initial (final) electron state.

Looking at Eq.(3) we see that the additional part does not contribute in
main logarithmic approximation and has not infrared divergence. Another
words, the born structure of the electron current tensor in the case of
longitudinally polarized electron is disturbed only in the
next--to--leading approximation due to radiation just by initial polarized
electron.

In general, the contribution into current tensor $L_{\mu\nu}$ due to
emission of $n$ collinear photons can be written as follows:
\begin{equation}\label{4 n photons}
L_{\mu\nu}^{(n)} =
\biggl(\frac{\alpha}{2\pi^2}\biggr)^n[I^{(n)}L_{\mu\nu}^B +
K^{(n)}i\lambda E_{\mu\nu}]\prod_{i=1}^n\frac{d^3k_i}{\omega_i} \ ,
\end{equation}
where the quantity $K^{(n)}$ equals to zero if (and only if) all $n$
collinear photons are emitted along the scattered (unpolarized) electron
momentum direction. The first term in the right side of Eq.(4) had been
obtained in [10] with the next--to--leading accuracy. Our goal is to find
also the second one with the same accuracy.

2.We use the covariant method of calculation and start from the general
expression for polarized current tensor which arises due to two hard
photon emission
\begin{equation}\label{5 tensor}
L_{\mu\nu}^{(2)} = \biggl(\frac{\alpha}{4\pi}\biggr)^2\frac{d^3k_1d^3k_2}
{\omega_1\omega_2}Sp(\hat p_2+m)Q_{\mu}^{\lambda\rho}(\hat
p_1+m)(1-\gamma_5\hat P)(Q_{\nu}^{\lambda\rho})^+ \ ,
\end{equation}
where $P$ is the polarization 4--vector of initial electron. The quantity
$Q_{\mu}^{\lambda\rho}$ reads
$$Q_{\mu}^{\lambda\rho} = \gamma_{\mu}\frac{\hat \Delta +m}{\Delta^2-m^2}
\gamma_{\rho}\frac{\hat p_1-\hat k_1+m}{-2p_1k_1}\gamma_{\lambda} +
\gamma_{\rho}\frac{\hat p_2-\hat k_2+m}{2p_2k_2}\gamma_{\mu}
\frac{\hat p_1-\hat k_1+m}{-2p_1k_1}\gamma_{\lambda} + $$
\begin{equation}\label{6 general spur}
\gamma_{\rho}\frac{\hat p_2=\hat k_2+m}{2p_2k_2}\gamma_{\lambda}
\frac{\hat\Sigma + m}{\Sigma^2-m^2}\gamma_{\mu} + (1\leftrightarrow 2) \
, \ \ \Delta = p_1-k_1-k_2 \ , \ \ \Sigma = p_2+k_1+k_2 \ .
\end{equation}

For the important case of the longitudinally polarized electron in the
frame of the choosen accuracy we can write the polarization vector in the
form
\begin{equation}\label{7 polarization vector}
P = \frac{\lambda}{m}\biggl(p_1-\frac{m^2k}{p_1k}\biggr) \ ,
\end{equation}
where $\lambda$ is the doubled electron helicity, and 4--vector $k$ has a
components $(\varepsilon_1,-\vec p_1), \ k^2 = m^2.$ It is easy to see that
$$ P^2 = -1+O(m^4/\varepsilon^4) \ , \ \ Pp_1 = 0 \ .$$
Note that for calculation in the leading approximation we can neglect with
the second term in the right side of Eq.(7) as it was done in [5].

There are four collinear regions in the case of double photon emission:
$(\vec k_1,\vec k_2 \parallel \vec p_1); \ (\vec k_1,\vec k_2 \parallel
\vec p_2); \ (\vec k_1 \parallel \vec p_1, \vec k_2 \parallel \vec p_2) $
and $(\vec k_1 \parallel \vec p_2, \vec k_2 \parallel \vec p_1).$ The
strightforward calculation in the region $(\vec k_1, \vec k_2 \parallel
\vec p_1$) when both hard collinear photons are emitted by the
initial--state polarized electron gives $$\frac{m^4}{4}I_{ii}^{(2)} =
\frac{1+y^2}{2x_1x_2\eta_1\eta_2} +
\frac{1}{d\eta_1}\biggl[-(1-x_2)+2y\Bigl(1-\frac{x_1}{x_2}\Bigr) +
\frac{1-x_1}{x_1x_2}((1-y)(x_1-x_2)-2y)\biggr] - \frac{y\eta_2}{d^2\eta_1}
+ $$
\begin{equation}\label{8 I region ii}
\frac{2}{d\eta_1^2}\Bigl(x_2+\frac{2y(1-x_1)}{x_2}\Bigr) -
\frac{4y}{d^2\eta_1} + \frac{(1-y)(2y+x_1x_2)}{x_1x_2d\eta_1\eta_2} +
\frac{4y}{d^2\eta_1}\Bigl(\frac{1}{\eta_1}+\frac{1}{\eta_2}\Bigr) +
(1\leftrightarrow 2)\ ,
\end{equation}

$$\frac{m^4}{4}K_{ii}^{(2)} =
\frac{2}{d\eta_1^2}\Bigl(1-x_2-x_1x_2+\frac{2yx_1^2}{x_2}\Bigr) +
\frac{1}{d\eta_1\eta_2}\Bigl(3-3y+2y^2+\frac{4x_2^2}{x_1}\Bigl) +
\frac{2}{d^2\eta_1}(3y-x_1^2-x_2^2) + $$
\begin{equation}\label{9 K region ii}
+\frac{2y\eta_2}{d^2\eta_1^2}
+
\frac{4}{d^2\eta_1}\Bigr(\frac{1}{\eta_1}+\frac{1}{\eta_2}\Bigr)[(1-y)^2-
x_1x_2] + (1\leftrightarrow 2)\ , \ \ y=1-x_1-x_2 \ .  \end{equation} When
writing Eqs.(8),(9) we used the following notations $$2p_1k_{1,2} =
m^2\eta_{1,2}\ , \ \ \Delta^2-m^2 = m^2d \ , \ \ x_{1,2}=
\frac{\omega_{1,2}}{\varepsilon_1} \ . $$

In the region $(\vec k_1, \vec k_2 \parallel \vec p_2)$ when both hard
colinear photons are emitted by the final--state unpolarized electron we
have
\begin{equation}\label{10 region ff}
K_{ff}^{(2)} = 0 \ , \ \ I_{ff}^{(2)} = I_{ii}^{(2)}(x_{1,2}\rightarrow
-y_{1,2}\ , \eta_{1,2}\rightarrow -\sigma_{1,2} \ , \ d\rightarrow \sigma
\ , \ y\rightarrow \eta=1+y_1+y_2) \ , \end{equation} where $$y_{1,2} =
\frac{\omega_{1,2}}{\varepsilon_2}\ , \ 2p_2k_{1,2} = m^2\sigma_{1,2} \ ,
\ \Sigma^2-m^2=m^2\sigma \ . $$

In accordance with the quasireal electron method [9] we can express the
electron current tensor in the region $(\vec k_1\parallel\vec p_1,
\vec k_2\parallel\vec p_2)$ as a production of the radiation probability
of the collinear photon with the energy $\omega_2$ by the scattered
electron (which is the coefficient at $L_{\mu\nu}^B$ in the right side of
Eq.(2) with $y=y_2$) and electron current tensor due to single photon
emission by the initial electron as given by Eq.(3) with $x=x_1$.
Therefore, the contribution of the regions $(\vec k_1\parallel\vec p_1,
\vec k_2\parallel\vec p_2)$ and $(\vec k_2\parallel\vec p_1,
\vec k_1\parallel\vec p_2)$ reads
\begin{equation}\label{11 regions if}
L_{\mu\nu}^{(2)if} =
\biggl(\frac{\alpha}{2\pi}\biggr)^2\Bigl[\frac{1+(1+y_2)^2}{y_2}\widetilde
L_0 -
\frac{2(1+y_2)}{y_2}\Bigr]\Bigl\{\Bigl[\frac{1+(1-x_1)^2}{x_1}L_0-\frac
{2(1-x_1)}{x_1}\Bigr]L_{\mu\nu}^B -
\end{equation}
$$2x_1i\lambda E_{\mu\nu}\Bigr\}dy_2dx_1 +(1\leftrightarrow 2)\ . $$

In order to derive the corresponding contributions in the regions
$(\vec k_1, \vec k_2\parallel\vec p_1) $ and $(\vec k_1, \vec k_2\parallel
\vec p_2) $ we have to perform the angular integration in Eq.(4) using Eqs.
(8) and (9). Moreover, we can also integrate over the energy fraction
$x_1 \ (y_1)$ in the region $(\vec k_1, \vec k_2\parallel\vec p_1) $
\ ($(\vec k_1, \vec k_2\parallel\vec p_2) $) at fixed value of the quantity
$x_1+x_2=1-y \ (y_1+y_2=\eta-1)$ because of 4--momentum of the heavy photon
which interacts with hadronic part of the amplitude depends on $1-y
\ (\eta-1)$ in this case.

The expressions (8) and (9) for $I_{ii}^{(2)}$ and $K_{ii}^{(2)}$ are
suitable for the calculations with a power accuracy (up to terms of the
order $m^2/\varepsilon_1^2$). But here we restrict ourselves with the
logarithmic accuracy and therefore can omitt terms proportional to
$1/d\eta_1\eta_2, \ 1/d^2\eta_1, \ 1/d^2\eta_1^2 $ and
$1/d^2\eta_1\eta_2$ in the right sides of Eqs.(8) and (9). In this
approximation the integration of the quantity $I_{ii}^{(2)}$ leads to
(see[10]) \begin{equation}\label{12 itegration I_ii}
\int\frac{d^3k_1d^3k_2}{\omega_1\omega_2}\frac{I_{ii}^{(2)}}{m^4} =
\pi^2\biggl[\frac{1}{2}L_0^2A(y,\delta) + L_0B(y,\delta)\biggr]dy \ ,
\end{equation}
\begin{equation}\label{13 leading part I for ii}
A=4\frac{1+y^2}{1-y}\ln\frac{1-y-\delta}{\delta}+(1+y)\ln y -2(1-y) \ ,
\end{equation}
\begin{equation}\label{14 next-to-leading part I for ii}
B=3(1-y) + \frac{3+y^2}{2(1-y})\ln^2y -\frac{2(1+y)^2}{1-y}\ln\frac
{1-y-\delta}{\delta} \ ,
\end{equation}
where $\delta<<1$ is the infrared cut for the energy fraction of each
photon. Analogously, the integration of the quantity $K_{ii}^{(2)}$ reads
\begin{equation}\label{15 K for ii}
\int\frac{d^3k_1d^3k_2}{\omega_1\omega_2}\frac{K_{ii}^{(2)}}{m^4} =
\pi^2L_0C(y,\delta)dy \ , \ C = 2(1-y)\biggl[2-\ln y - 2\ln\frac
{1-y}{\delta}\biggr]dy \ .
\end{equation}
By using the Eqs.(12),(15) together with Eq.(4) we obtain
\begin{equation}\label{16 total ii}
L_{\mu\nu}^{(2)ii} =
\biggl(\frac{\alpha}{2\pi}\biggr)^2\Bigl[\Bigr(\frac{1}{2}L_0^2A(y,\delta)
+L_0B(y,\delta)\Bigr)L_{\mu\nu}^B + C(y,\delta)L_0i\lambda
E_{\mu\nu}\Bigr]dy
\end{equation}
for the contribution of the region $(\vec k_1, \vec k_2 \parallel\vec p_1)$
into the current tensor of longitudinally polarized electron. In some
applications the quantyity $y$ kept fixed (for example, for calculation of
the tagged photon cross--sections). In this case we can write
$\ln((1-y)/\delta)$ instead of $\ln((1-y-\delta)/\delta)$ in expressions
for the quantities $A$ and $B$.

The corresponding contribution of
the region $(\vec k_1, \vec k_2 \parallel \vec p_2)$ can be written as
follows
\begin{equation}\label{17 total for ff} L_{\mu\nu}^{(2)ff} =
\biggl(\frac{\alpha}{2\pi}\biggr)^2\Bigl[(\frac{1}{2}\widetilde
L_0^2\widetilde A(\eta,\delta') + \widetilde
L_0\widetilde B(\eta,\delta')\Bigr]L_{\mu\nu}^B d\eta \ ,  \delta' =
\frac{\delta\varepsilon_1}{\varepsilon_2}\ ,
\end{equation}
where
\begin{equation}\label{18 leading part for ff}
\widetilde A = 4\frac{1+\eta^2}{1-\eta}\ln\frac{\eta-1-\delta'}{\delta'}
-(1+\eta)\ln\eta -2(\eta-1) \ ,
\end{equation}
\begin{equation}\label{19 next-to-leading part for ff}
\widetilde B = 3(\eta-1) + \frac{3+\eta^2}{2(\eta-1)}\ln^2\eta
-2\frac{(1+\eta)^2}{\eta-1}\ln\frac{\eta-1-\delta'}{\delta'} \ .
\end{equation}
Note that the quantities $\widetilde A$ and $\widetilde B$ can be
reconstructed from the quantities $A$ and $B$ by the rule
$$\widetilde A(\eta,\delta') = - A(\eta,-\delta') \ , \ \ \widetilde
B(\eta,\delta) = - B(\eta,-\delta') \ .$$

As we saw above (Eq.(3)) the additional part to the Born structure of
polarized electron current tensor due to single collinear photon emission
has neither colinear (does not contain large logarithm) nor infrared (is
finite in the limit $x$ go to zero) singularities. But these singularities
appear in the corresponding contribution due to double collinear photon
emission (Eqs.(11),(16)). Nevertheless, the additional part never
conribute in the leading aproximation.

The infrared parameter $\delta$ must eliminate in any physical application
if photons are unobserved. Such elimination takes place because of
contributions due to double virtual and soft photon emission as well as
virtual and soft correction to single hard photon emission. The last
contribution have been consider recenty [5] inside approximation $m^2=0$
which describes the large--angle photon radiation. If we put $m^2=0$ in
our calculations than we will leave only with the born--like structure in
Eqs.(3),(11) and (16). Moreover, quantities $B$ and $\widetilde B$ in
Eqs.(16) and (17) will be changed in this approximation. We see
consequently, that it needs to keep finite the electron mass to be correct
inside next--to--leading approximation in any physical application with
unobserved photons (for example in classical deep inelastic scattering).
We conclude therefore that the results of work [5] have to be improved for
such kind applications.

3.Let us consider double hard photon emission in semicollinear regions
$\vec k_1 \parallel \vec p_1$ or $\vec p_2$ and $\vec k_2$ is arbitrary.
In this situation we can use the quasireal electron method for
longitudinally polarized initial electron [9]. In accordance with this
method the contribution of the region $\vec k_1 \parallel \vec p_2$ into
electron current tensor is defined by its born--like structure
$L_{\mu\nu}^{^{\gamma}}$ as follows
\begin{equation}\label{20 semicollinear f}
L_{\mu\nu}(\vec k_1\parallel\vec p_2) =
\frac{\alpha^2}{8\pi^3}\frac{d^3k_2}{\omega_2}\frac{dy_1}{1+y_1}
\Bigl[\frac{1+(1+y_1)^2}{y_1}\widetilde L_0 - \frac{2(1+y_1)}{y_1}\Bigr]
L_{\mu\nu}^{^{\gamma}}(p_1,p_2(1+y_1),k_2) \ ,
\end{equation}
where for large--angle emission tensor $L_{\mu\nu}^{^{\gamma}}$ we can use
the approximation $m^2=0$ [3,5,11]
\begin{equation}\label{21 born large angle}
L_{\mu\nu}^{^{\gamma}}(p_1,p_2,k_2) = 4(B_{\mu\nu} + i\lambda
E_{\mu\nu}^{^{\gamma}}) \ ,
\end{equation}
$$B_{\mu\nu} = \frac{1}{st}[(s+u)^2+(t+u)^2]\widetilde g_{\mu\nu} +
\frac{4q^2}{st}(\tilde p_{1\mu}\tilde p_{2\nu} + \tilde p_{1\nu}p_{2\mu})
\ , $$
$$E_{\mu\nu}^{^{\gamma}} =
\frac{2\epsilon_{\mu\nu\rho\sigma}}{st}[(u+t)p_{1\rho}q_{\sigma}+(u+s)
p_{2\rho}q_{\sigma}] \ , \ \ \widetilde g_{\mu\nu} =
g_{\mu\nu}-\frac{q_{\mu}q_{\nu}}{q^2} \ , $$
$$\tilde p_{\mu} = p - \frac{(pq)q_{\mu}}{q^2}\ , \ u=-2p_1p_2\ , \
s=2p_2k_2 \ , \ t=-2p_1k_2 \ , \ q=p_2+k_2-p_1 \ . $$

As above, the emission of collinear photon by the initial electron
disturbs the Born structure of the electron current tensor just in the
same manner as it done in Eq.(11)
\begin{equation}\label{22 semicollinear i}
L_{\mu\nu}(\vec k_1\parallel\vec p_1) =
\frac{\alpha^2}{8\pi^3}\frac{d^3k_2}{\omega_2}\frac{dx_1}{1-x_1}\biggl\{
\bigl[\frac{1+(1-x_1)^2}{x_1} L_0 - \frac{2(1-x_1)}{x_1}\bigr]
L_{\mu\nu}^{^{\gamma}}(p_1(1-x_1),p_2,k_2) -
\end{equation}
$$2x_1i\lambda E_{\mu\nu}^{^{\gamma}}(p_1(1-x_1),p_2,k_2)\biggr\}$$

Formulae (20) and (22) are derived by us independently on quasireal
electron method starting from the general expression for current tensor as
given by Eqs.(5), (6) and (7).

When calculating the radiative corrections to polarized DIS cross--section
we have to integrate over all phase space of photons. At this case the
angular cut parameter $\theta_0$ is unphysical and must vanish in the sum
of contributions of double collinear and semicollinear regions. At taken
accuracy this fact goes to cancellation terms of the type
$L_0\ln\theta_0^2$, and that can be verifyied by separation of
$\ln\theta_0^2$ at integration of $L_{\mu\nu}^{^{\gamma}}(p_1(1-x_1),p_2,k_2)$
in the limit $\vec k_2\parallel \vec p_1$:
\begin{equation}\label{23 elimination theta}
\int\frac{d^3k_2}{\omega_2}L_{\mu\nu}^{^{\gamma}}(p_1(1-x_1),p_2,k_2\approx
x_2p_1) = -
2\pi\ln\theta_0^2dx_2\frac{y^2+(1-x_1)^2}{(1-x_1)x_2}L_{\mu\nu}^B \ .
\end{equation}
Taking into account that at fixed $x_1+x_2=1-y$
\begin{equation}\label{24}
\int dx_1dx_2\frac{[1+(1-x_1)^2][y^2+(1-x_1)^2]}{x_1x_2(1-x_1)^2} =
A(y,\delta)dy \ ,
\end{equation}
we carry conviction that the terms of the type $L_0\ln\theta_0^2$ indeed
vanish in the sum of contributions due to colinear kinematics and
semicollinear one. Analogous cancellation take place of course for
radiation in the final state.

Note in conclusion that the electron current tensor has an universal
character.  It can be used for calculation of cross--sections in different
processes including most interesting DIS and $e^+e^-$--annihilation into
hadrons.  To obtain the corresponding cross--sections we have to multiply
the electron current tensor by the hadron one. Just hadron tensor carry
important information about hadronic structure and fragmentation
functions [12], and the study of radiative corrections to electron current
tensor is necessary for interpretation experimental data in terms of
these hadronic functions.

Authors thank A.B.Arbuzov and I.V. Akushevich for discussion. This work
supported in part (N.P.M.) by INTAS grant 93--1867 ext and Ukranian DFFD
by grant N 24/379.  \\

\vspace{0.5cm}
\hspace{1.0cm}
{\bf {\large {References}}} \\

\vspace{0.5 cm}
\begin{enumerate}
\item SMC., D. Adams et al., Phys.Rev. {\bf D 56} (1997) 5330.
\item HERMES, K. Acherstaff et al., Phys.Lett. {\bf B 404} (1997) 383.
\item T.V. Kukhto and N.M. Shumeiko, Nucl.Phys.{\bf B 219} (1983) 412.
\item I.V. Akushevich and N.M. Shumeiko, J.Phys. {\bf G 20} (1994) 513.
\item I.V. Akushevich, A.B. Arbuzov and E.A. Kuraev, Phys.Lett. {\bf B
432} (1998) 222.
\item A.B. Arbuzov et al., Nucl.Phys. {\bf B 485} (1997)
457; Phys.Lett.{\bf B 399} (1997) 312; N.P. Merenkov, JETP {\bf 112}
(1997) 400.  \item H. Anlauf, A.B. Arbuzov, E.A. Kuraev and N.P. Merenkov,
HEP--PH/9711333.
\item A.B. Arbuzov, E.A. Kuraev, N.P. Merenkov and L. Trentadue, (to be
published in JHEP).
\item V.N. Baier, V.S. Fadin V.A. Khoze, Nucl.Phys. {\bf B 65} (1973) 381;
V.N.Baier, V.S. Fadin V.A. Khoze and E.A. Kuraev, Phys.Rep. {\bf 78}
(1981) 293.
\item N.P. Merenkov, Sov.J.Nucl.Phys. {\bf 48} (1988) 1073.
\item E.A. Kuraev, N.P. Merenkov and V.S. Fadin, Yad.Fiz. {\bf 45} (1987)
782.
\item X. Ji, Phys.Rev. {\bf D 49} (1994) 114; R.L. Jaffe and X. Ji,
Phys.Rev.Lett. {\bf 67} (1991) 552.
\end{enumerate}

\end{document}